\documentclass{article}
\usepackage[utf8]{inputenc} % set input encoding (not needed with XeLaTeX)

\usepackage{geometry} % to change the page dimensions
\usepackage{graphicx} % support the \includegraphics command and options

\usepackage{booktabs} % for much better looking tables
\usepackage{array} % for better arrays (eg matrices) in maths
\usepackage{paralist} % very flexible & customisable lists (eg. enumerate/itemize, etc.)
\usepackage{verbatim} % adds environment for commenting out blocks of text & for better verbatim
\usepackage{subfig} % make it possible to include more than one captioned figure/table in a single float
\usepackage{amsfonts,amssymb}
\usepackage{amsmath}
\usepackage{latexsym}
\usepackage{esint}
\usepackage{graphicx}
\usepackage{float}
\usepackage{color}
\usepackage{setspace}
\usepackage{booktabs}
\usepackage{multirow}
\usepackage[colorinlistoftodos]{todonotes}
\usepackage{amsfonts}
\usepackage{fancyhdr} % This should be set AFTER setting up the page geometry
\usepackage[nottoc,notlof,notlot]{tocbibind} % Put the bibliography in the ToC
\usepackage[titles,subfigure]{tocloft} % Alter the style of the Table of Contents

\title{Fermi Surface Geometry}
\author{Elena Derunova$^1$$^{\dagger}$, Jacob Gayles$^2$, Yan Sun$^3$, Michael W. Gaultois$^4$, Mazhar N. Ali$^1$*}

\begin{document}

\maketitle
$^1$ Max Plank Institute of Microstructure Physics, Halle, Germany.\newline %1
$^2$ Department of Physics, University of South Florida, Tampa, Florida 33620, USA \newline %2
$^3$ Max Planck Institute for Chemical Physics of Solids, Dresden, Germany \newline %3
$^4$ Leverhulme Research Center for Functional Materials Design, The Materials Innovation Factory, Department of Chemistry, University of Liverpool, Liverpool, UK \newline%3

\begin{abstract}
\indent{}\textbf{Motivated by the famous and pioneering mathematical works by Perelman \cite{Perelman_2008}, Hamilton \cite{hamilton1982}, and Thurston \cite{thurston_three_nodate}, we introduce the concept of using modern geometrical mathematical classifications of multi-dimensional manifolds to characterize electronic structures and predict non-trivial electron transport phenomena. Here we develop the Fermi Surface Geometry Effect (FSGE), using the concepts of tangent bundles and Gaussian curvature as an invariant. We develop an index, $\mathbb{H}_F$, for describing the the "hyperbolicity" of the Fermi Surface (FS) and show a universal correlation (R$^2$ = 0.97) with the experimentally measured intrinsic anomalous Hall effect of 16 different compounds spanning a wide variety of crystal, chemical, and electronic structure families, including where current methods have struggled. This work lays the foundation for developing a complete theory of geometrical understanding of electronic (and by extension magnonic and phononic) structure manifolds, beginning with Fermi surfaces. In analogy to the broad impact of topological physics, the concepts begun here will have far reaching consequences and lead to a paradigm shift in the understanding of electron transport, moving it to include \textit{geometrical properties} of the E vs k manifold as well as topological properties.}  
%An apparent maximum value at $\mathbb{H}_F$ = 1, of ~1570 $\frac{\hbar}{e}(\Omega cm)^{-1}$ was determined for materials with an FS created by bands belonging to a single elementary band representation (EBR); materials with multi-EBR FS's can, and do, break this limit as evidenced by CrPt$_3$ and Co$_2$MnAl. 
\end{abstract}

\newpage\section{Introduction}
\onehalfspacing
\indent{}While the mathematical concept of topology originally arose as a purely theoretical discipline, topological physics exploded onto the scene after the experimental realization of Topological Insulators (TI) \cite{zhang2009topological,Konig02112007,cercellier2007evidence} in 2007. This opened doors for mathematics mixed with material science to become an important fundamental and technological topic for researching novel physical phenomena as well as technologically relevant properties \cite{Konig02112007, ohno2016spintronics,RevModPhys.80.1517}. The idea of using numbers for the classifications of energy manifolds in crystalline solids, i.e. electronic bands in energy-momentum space (E vs k), has been used to successfully predict protected electronic states and properties including the quantum Hall effects \cite{liu2011quantum,kane2005quantum}. There have also been attempts to use topology to characterize other types of experimentally observed transport phenomena (e.g. the non-quantized anomalous Hall effect (AHE), spin Hall effect(SHE), skyrmion/domain wall dynamics and superconductivity) based on the non-quantized properties of a gauge generated field on the electronic band which is known as the Berry phase \cite{Xiao2010}. This Berry phase approach works relatively well for predicting the intrinsic AHE and SHE (where carriers acquire a velocity component orthogonal to an applied electric field without an externally applied magnetic field), but has had accuracy issues with compounds like Ni \cite{theo_ni_co_fe}. The topological theories of electronic band structures for strongly correlated phenomena like superconductivity and charge density wave formation, however, are not as successfully used as macro models like Ginzburg-Landau and Mott-Hubbard theory \cite{ginzburg,CYROT1977141}. At the same time topology, as a broad discipline, has continued to move forward and the use of a number as a classifying invariant has grown to include the use of a geometrical structure playing the same role.    

\indent{}An electronic energy band is, fundamentally, a 3 dimensional orientable, closed manifold; it encompasses a compact topological space that has a consistent choice of a local orientated (left hand/right hand) Cartesian coordinate frame defined over the whole reciprocal (k) space, and thus has no boundaries. For these types of manifolds, the geometrical structures which are possible to exist were first theorized by Thurston \cite{thurston_three_nodate} and proved recently by Perelman \cite{Perelman_2008}. This famously led to the proof of the Poincare Conjecture (one of Millennium Prize Problems and one of the most important questions in topology). According to his work, a 3D manifold, like an electronic energy band, can be split into regions, where each region can be classified into one of the 8 Thurston geometrical classes (schematically described in Figure 1). Since the electronic structure fundamentally governs intrinsic (not scattering driven) electron transport phenomena in crystalline materials, the geometrical classes (or combinations of them as well as the boundaries between them) must also correspond to intrinsic transport phenomena. Similar to how topology has been used to classify band structures and relate them to transport phenomena, a \textit{geometrical} understanding of band structures, with classification and relation to transport, can be carried out. In doing so, one of most fundamental results in modern mathematics may be found to have direct practical application – not only to condensed matter physics and material science, but also to device technologies. However, developing the complete theory of Electronic Structure Geometry is highly challenging and beyond the authors’ current abilities and the scope of this work.  

\indent{}For now, we simplify this task by working with 2D cross-sections of the 3D electronic structure, looking at constant energy surfaces of the 3D manifold. In particular, we take the cross-section (of E vs k) at the Fermi energy (corresponding to the highest occupied energy states i.e. the highest occupied molecular orbital), which is also known as the Fermi Surface (FS). The details of the FS determine electron transport properties in metals and semimetals and, in the assumption that it is a differentiable surface, the geometrical classifications simplify from eight to just three types: elliptic, hyperbolic, and Euclidean (Figure 1). In analogy to what Perelman showed for a 3D manifold, the 2D Fermi surface can also be understood as a composition of local regions of these different types. The behavior of electrons on the elliptic regions (similar to electrons on a sphere) are already well described via conventional Fermi-liquid theory. Non-Fermi-liquid behavior of electrons, including topological transport like Berry phase driven phenomena, must be related to the remaining two geometric types and thus studying them can potentially lead us to understanding and predicting anomalous transport behavior.

\indent{}In this work we focus in particular on the hyperbolic geometrical class and investigate its relationship to the intrinsic anomalous and spin Hall effects. We use the concepts of tangent bundles and Gaussian curvature to develop the metric $\mathbb{H}_F$ for measuring the hyperbolicity of the Fermi surface (how much of the FS is hyperbolic and oriented in a way relevant to the Hall effect in question) and show that it correlates extremely well with experimentally measured values of intrinsic anomalous Hall conductivity (AHC) (R$^2$ = 0.97). We show this for 16 different real materials ranging from conventional ferromagnets to Weyl semimetals, including cases like Ni and Co$_2$MnAl, where the Berry phase approach is not accurate. This Fermi Surface Geometry Effect (FSGE) also works consistently with the recent formulation of topological quantum chemistry \cite{Bradlyn2017}; we found that 14 of the compounds have FSs generated from bands with a single elementary band representation (EBR) and that the limit of the AHC for a single EBR FS is ~1570 $\frac{\hbar}{e}(\Omega cm)^{-1}$. Two of the materials examined here, CrPt$_3$ and Co$_2$MnAl, have multi-EBR Fermi surfaces and subsequently break the apparent AHC limit. The strategy to maximize the intrinsic AHC, therefore, is to find materials with highly hyperbolic, multi-EBR FS’s. We also find that the FSGE matches predictions of the spin Hall conductivity (SHC) for Pt, Beta-W, and TaGa$_3$. The $\mathbb{H}_F$ index enables a cheap and rapid computational prediction of AHE/SHE materials and can be implemented with existing density functional theory (DFT) methods and databases.     

\indent{}The profound result is that, even with the 3D to 2D submanifold simplification, we are able to apply modern geometrical mathematical concepts to correlate geometrical structure with non-trivial transport phenomena in a highly diverse set of compounds where conventional methods have struggled. This illustrates the importance of developing a complete theory of \textit{geometrical} understanding of Fermi surfaces and, eventually, electronic band structures, as a natural extension to the last decade of development in the topological understanding of band structures. This use of the results from the Poincare Conjecture proof will have far reaching consequences in the field of condensed matter physics and, like topological theory, will lead to a fundamental shift in the understanding of electron transport, moving it to include geometrical properties of the E vs k manifold.  

\section{Results and discussions}

\indent{}To begin with, the Bloch theory of electrons traveling in a periodic potential created by the crystalline lattice allows for the consideration of electron dynamics to be shifted from real space to momentum space (i.e. reciprocal space or k-space) and the Brillouin zone (BZ; the primitive cell in the reciprocal lattice). Electronic energy bands are created by the distinct energy levels, $\epsilon_n$, of the Bloch functions describing the extended orbitals of the lattice, $\psi_n(k,X) = (e^{ikna}) u_n(k,X) $, where  $k=(k_x,k_y,k_z)$ is the index, or momentum, $a=(a_1,a_2,a_3)$ is the periodicity in the BZ, $u_n$ is the basis function of the orbital, and $X=(x,y,z)$ is the coordinate of the electron in real space. In the BZ, the $\epsilon=\epsilon_n( k_x,k_y,k_z)$ is the set of energy bands for all orbitals in the lattice and is known as the electronic bandstructure. Then, the dynamics of an electron traveling in the crystal potential is described by the following equations \cite{AshcroftMermin}:

% semiclassical transport
\begin{equation}\label{semiclas_k}
\hbar  \frac{dk_a}{dt}=eE(x)+eH_{ab}(k)\frac{dx^b}{dt}
\end{equation}

\begin{equation}\label{semiclas_x}
\frac{dx^a}{dt}=\frac{1}{\hbar}\nabla_a \epsilon(k)
\end{equation}

\indent{}where  E(x) is an applied electric field, H(x) is an applied magnetic field. As it is well known from Fermi Liquid Theory, the many$-$electron dynamics in the crystal potential can be described by the same equations, but applied to "`quasi-particles"' with renormalized dynamical properties as compared to fermions in a Fermi gas \cite{AshcroftMermin}. One can derive from the above equations that without an applied field, a quasi-particle moves along the normal vector $v_F=(\partial \epsilon/\partial k_x,\partial \epsilon/\partial k_y\partial \epsilon/\partial k_z)$ to the constant energy surface (i.e. group velocity) and under an applied electric field the quasi-particle moves from one constant energy surface in k-space to another in the direction of the field, which is to say that there is a $\delta k$ along a 1D cross-section of the full 3D band. To make equation (2) useable, the function $\epsilon(k)$ should be differentiable; in the quasi-particle approximation, particularly, it should be differentiable up to 2nd order (i.e., the function $\epsilon(k)$ should be smooth). 

\indent{}The band structures for real compounds, however, have also shown the presence of \textit{singular} points from intersections of locally linearly dispersive bands, also known as Dirac or Weyl points (depending on degeneracy). In these cases, $\epsilon(k)$ is non-smooth, meaning the derivative $\partial\epsilon/\partial k_a$ is not well defined. Thus equation (2) cannot be applied and the Fermi liquid quasi-particle approximation is no longer valid. A plethora of work in the last decade has dealt with the ramifications of this, with Dirac/Weyl points resulting in the creation of "`Dirac/Weyl quasi-particles"' and the AHE/SHE due to uncompensated Berry curvature arising around a Dirac type degeneracy in the bulk band structure \cite{gradhand2012first}. In this case, Berry curvature, is used to quantify the effect of these special points in the 3D bandstructure \cite{gradhand2012first, PhysRevB.77.165117, RevModPhys.87.1213, PhysRevLett.100.096401}, defined as:

\begin{equation}\label{berry_one}
\Omega_n(k)=\nabla_k \times <u_n(k,X),i\nabla_k, u_n(k,X)>
\end{equation}

\indent{}It gives information about the antisymmetric behavior of the changing energy around the degenerate point and, particularly, the rotational component of the group velocity vector field over reciprocal space. To quantify its effect on the AHE/SHE, perturbations like SOC, which result in recovering the smoothness of the corresponding Dirac points, are manifested proportionally to the energy changes that have been brought into the bandstructure via the perturbation: 

\begin{equation}\label{kubo_ahc}
\sigma_{xy}^z=e^2\hbar\int_{BZ} \frac{1}{2\pi}\Omega_{xy,n}^z f_n(k) dk^3
\end{equation} 

\indent{}Where $f_n(k)$ is the Fermi distribution function. The calculation of AHC/SHC is carried out in practice using the Kubo formalism based on inter-band exchange resulting in Berry curvature\cite{gradhand2012first}: 
\begin{equation}
\Omega_{xy,n}^z(k)=-\hbar^2 \sum_{m\ne n}\frac{Im[<\psi_{nk} \mid v_x \mid \psi_{mk}>\times <\psi_{mk}\mid v_y \mid \psi_{nk}>]}{(\varepsilon_{nk} - \varepsilon_{mk})^2}
\end{equation}

\indent{}The Berry curvature, $\Omega_{k_i,k_j}$, is modulated by the SOC magnitude and diverges around Dirac/Weyl points appearing in the $(k_i,k_j)$ plane in the 3D bandstructure of the unperturbed electron state \cite{Derunovaeaav8575}, resulting in large AHE/SHEs. 

\indent{}Similar to how SOC is handled as a perturbation in the electron's Hamiltonian, the local geometry of the FS and the paths on the FS which a quasi-particle travels (FSP - Fermi Surface path) can be handled as a perturbation to the electron's momentum $\delta k$, resulting in observable transport properties. Dirac/Weyl points in the 3D bandstructure result in non-smooth regions of the FS and correspondingly irregular FSPs, meaning there is at least one direction where the curve $Ef(k_i(s),k_j(s), k_l)$ cannot be parametrized as $Ef(k_i(k_l),k_j(k_l), k_l)$. For example, if a Dirac-type bandstructure exists as shown in supplement Figure S1 with a Dirac crossing along the $k_x$ direction, an irregular FSP is generated by that crossing in the ($k_x$,$k_y$) plane (shown in purple). By commutation, these types of irregular points and paths should correlate with large AHE/SHEs. However there are more types of FS geometries and FS paths aside from just non-smooth and irregular ones arising from Dirac/Weyl points in the bandstructure, which may contribute to the transverse velocity of an electron (amongst other transport phenomena). 

\indent{}As shown in Figure 1 and discussed earlier, the 8 Thurston geometries classifying 3D bandstructures can result in 3 classes of smooth 2D Fermi Surfaces as well as non-smooth FSs. Correspondingly, several types of regular and irregular Fermi Surface paths can be generated from more than just non-smooth FSs. This raises the question: do other geometrical classes also contribute to anomalous transport behavior like AHE/SHE? If so, using a quantification method like the Kubo formalism, which essentially only considers one class of FS geometry and FSP, may incorrectly estimate the extent of anomalous transport governed the FS. As shown in Figure 1, the different FS geometrical classes correspond to particular classes of FSPs; locally elliptic FS’s result only in locally regular closed FSPs, locally Euclidean FS’s only result in locally regular open FSPs, and hyperbolic FS’s can result \textit{all classes of FSPs}. Hence we focus our study on quantifying the local hyperbolic FS regions. A full analysis of non-smooth regions and region boundaries of the FS is also an important direction towards developing a complete theory, but is also beyond the scope of this work. 

\indent{}In order to identify and quantify the local hyperbolic regions and take into account FSPs beyond only those considered by the Kubo formalism we use the concept of \textit{tangent bundles}. The tangent bundle, denoted as $TE_F$, is the set of the tangent planes at every point on the surface. The tangent plane at the point $K$, denoted as $T_KE_F$, is a plane spanned by the tangent vectors for \textit{all possible FSPs passing the point K}, schematically shown as a green plane in the Figure 2a on an example spherical FS. This plane implicitly holds information about $E_n (k+\delta k)$ which is important for understanding quasi-particle dynamics in the electric field. This information can be recovered using the First and the Second Fundamental Forms \cite{toponogov2006differential}:

\begin{equation}\label{fund_forms1}
I=
 \begin{pmatrix}
 g_{11} & g_{12}  \\
 g_{21} &  g_{22} 
\end{pmatrix},
g_{ij}=<\delta k_i, \delta k_j> 
\end{equation}

\indent{}Where $\delta k_i$ is the tangent vector to the FS in the $i$ direction. The scalar product, $g_{ij}$, in the tangent plane $T_KE_F$ at every point $K$ defines a Riemannian metric on the Fermi surface. Hence the Fermi surface can be considered as a 2D Riemannian manifold \cite{jost2005riemannian}. 
\begin{equation}\label{fund_forms2}
II=
 \begin{pmatrix}
L & M  \\
M & N  
\end{pmatrix},
k_l=L\frac{{k_i}^2}{2}+Mk_ik_j+N\frac{{k_j}^2}{2}+...
\end{equation}

\indent{}Where $k_l=f(k_i,k_j)$, a parametrization of surface. The Gaussian curvature is then defined by the ratio $K=detII/detI$ and used as classifying invariant of the local geometry of the surface in the following way: $K > 0$ corresponds to elliptic; $K = 0$ to Euclidean, and $K < 0$ corresponds to hyperbolic ( fig 2 b,c,d) \cite{poincare1908, uni_th}. In fact Euclidean geometry is particularly hard to distinguish from hyperbolic or elliptic, since numerical simulation of the FS always gives a small computation error which results in a very small positive or negative K; hence numerical determination of Euclidean regions are heavily dependent on practical tolerance factors. Every type of geometry also has a different dimensionality of the set of shared points between the FS and the tangent plane: a point, lines, or plane, is schematically shown in the fig.2 b,c,d in black color on the green square representing $T_KE_F$, and the solid line is the intersection of the FS and $T_KE_F$. %and the dashed line is the projection of the Fermi surface to the tangent plane $T_KE_F$ at the point K. 

\indent{}Let us assume that the \textit{tangent plane is parallel to the AHE plane}. The green arrows in the figure then represent the in-plane component of the group velocity $v_F^{xy}$, i.e. the projection of $v_F$ to the AHE plane. As one can see for the elliptic case, the electron can have a $v_F^{xy}$ in any direction on $T_KE_F$ but for the Euclidean case there is no in-plane component of the $v_F$ and thus dynamics may happen only out of plane. For the hyperbolic case, however, $T_KE_F$ is split into two regions; $v_F^{xy}$ pointing towards $K$, where quasi-particle doesn't leave the tangent plane unless it moves into the other region, $v_F^{xy}$ pointing away from K. Around an infinitesimal neighborhood of K, $v_F^{xy} $ is demanded to be in a single direction, for example, the $y$ direction. This $v_F^{xy}$ locking of one particular direction is manifested as an AHE/SHE, shown schematically in the fig 2e. Thus we expect that the total amount of hyperbolic points at the FS may be roughly correlated with the magnitude of the intrinsic AHE/SHE. 

\indent{}To explore this, we plot the distribution of hyperbolic regions on the FS for a few well known AHE/SHE compounds by carrying out the Gaussian curvature evaluation for every point on the k-mesh (see supplement for details and the lack of effect of k-mesh density beyond a low, cut-off value). The FS of Fe (Figure 2F), well known for its large intrinsic AHE of $~$1000 $\frac{\hbar}{e}(\Omega cm)^{-1}$ (experimentally determined, \cite{miyasato_crossover_2007}), clearly shows a preponderance of hyperbolic points compared with non-hyperbolic points comprising its FS. There is also a clear majority of negatively contributing hyperbolic points compared with positively contributing hyperbolic points (decided by the sign of the local curvature of the electronic band at the Brillouin zone boundary (BZB); see supplement). Co$_2$FeSi, another known AHE compound, but with a much lower experimentally measured intrinsic AHE ($~$200 $\frac{\hbar}{e}(\Omega cm)^{-1}$) (ref), also has a many hyperbolic points comprising its FS, but is highly compensated: positive and negative contributions are nearly equal. In analogy to Berry curvature compensation, explained in the work \cite{Xiao2010}, oppositely signed regions contribute with opposite sign to the AHE, thus for Fe the large magnitude of AHE can be the result of an uncompensated hyperbolic FS. Also in Figure 2F, the FS of Pt, having SHC around 2000 $\frac{\hbar}{e}(\Omega cm)^{-1}$, clearly shows a majority of hyperbolic points, whereas TaGa$_3$, with much lower SHC (See figure S4 in the supplement), has a much lower percentage of hyperbolic points comprising its FS.

\indent{}It is also important to consider the case where $T_KE_F$ is not exactly parallel but instead \textit{tilted with respect to the AHE plane}; the tangent plane will still have similar behavior as described above, but the magnitude of the $v_F^{xy}$ should now be proportional to the angle $\phi$ between $T_KE_F$ and AHE plane. Then to estimate contribution of the hyperbolic points we need to consider the projection of the unit vector of the tangent plane in the direction of the expected AHE current in the AHE plane (which we denote $F_{Hall}$) as shown in the Figure 3a.

\indent{}To more rigorously quantify the correlation between hyperbolic points on the FS and AHE/SHE we introduce the index of “hyperbolicity” of the FS, which we denote by $\mathbb{H}_F$ and define as the following:
% and thus there are more points which should also contribute to the AHE/SHE
%formula, Eq. N
\begin{equation}\label{H}
\mathbb{H}_F=\frac{\sum_{FS}I_nF_{Hall}(K\  is \  hyperbolic)}{\sum_{FS}F_{Hall} (K\ is\  arbitrary)}
\end{equation} 

\indent{}Where  $F_{Hall}$, is the projection of the unit vector, $e$ in Figure 3a is the vector parallel to $T_KE_F$ at point K, in the direction of analyzed Hall current (orthogonal to the applied current) and $I_n$ is the sign of the $\partial^2\epsilon_n/\partial^2k$ at the BZ boundary in the Hall current direction. As defined, $\mathbb{H}_F$ can have a maximum of “1” which means that the entire FS would be hyperbolic, except for the points where the tangent plane is orthogonal to the AHE plane. %with $F_Hall$ with all tangent planes at all points of the FS aligned with the Hall direction: an impractical scenario. 	 - not the case here 

\indent{}We performed unperturbed DFT calculations of 16 compounds for which intrinsic AHC values were rigorously experimentally determined\cite{liu_giant_2018, miyasato_crossover_2007, nayak_large_2016, kubler_weyl_2017,zeng_linear_2006,Fang92,CoFeSiAl, tung_high_2013}, covering a variety of structural families (perovskites, Heuslars, Kagome lattices, FCC lattices, etc.) and topological classes (Dirac/Weyl/Trivial metals and semimetals). We compared those experimental AHC values to our calculated $\mathbb{H}_F$ (taking care to align the directions of calculation with the directions of measurement for each material in the various experiments). Since, for now, the $\mathbb{H}_F$ parameter is defined for a 2D conductance, we made a comparison of the calculated $\mathbb{H}_F$ with the measured values for thin films or layered structures where the contribution of the third dimension to the total effect is relatively weak. 

\indent{}The result, shown in Figure 3c, shows an extraordinary linear correlation of the hyperbolicity of the FS with the experimentally measured AHC values of all compounds, regardless of structural family or topological class, with an $R^2$ value of 0.97. Importantly, we also plot the experimentally measured intrinsic AHC values versus the calculated AHC values using the Kubo formalism (based on Berry curvature \cite{liu_giant_2018, zeng_linear_2006,kubler_weyl_2017,Fang92,CoFeSiAl,theo_ni_co_fe, tung_high_2013}) for the same compounds (Figure 3d). The $R^2$ drops down to only 0.52, with a few exceptionally inaccurate cases like Co$_2$MnAl or Ni, where the error is large and the reason is still under investigation \cite{NiT}. Even without taking those two compounds into account, the $R^2$ from the Kubo calculated AHCs only rises to 0.87; significantly worse than the $\mathbb{H}_F$-dependence. We made a similar calculation for SHE compounds, but due to a paucity of experimental data, we are forced to plot the comparison of the Kubo predicted SHC values for Pt, W$_3$W\cite{gradhand2012first,Derunovaeaav8575}  and TaGa$_3$  at different $E_F$-levels against their corresponding $\mathbb{H}_F$-values in the Figure 3b. This graph also shows a strong correlation of the hyperbolicity also with the Kubo calculated SHCs with an $R^2$ of about 0.95, implying that the Kubo approach and the hyperbolicity match well in the case of highly uncompensated FS’s \cite{6516040} (Figure 2 illustrated the uncompensated FSs of these SHE compounds); a direction of future investigation.   

\indent{}From the correlation in Figure 3c, it can be seen that in the limit of $\mathbb{H}_F$ = 1, the intrinsic AHC is expected to reach a maximum value of ~1570 $\frac{\hbar}{e}(\Omega cm)^{-1}$. However there are two compounds (Co$_2$MnAl and CrPt$_3$) that have $\mathbb{H}_F$ and intrinsic AHCs greater than these maxima. While at first this appears to be an inconsistency, the limit on $\mathbb{H}_F$ can be broken if we take into account the EBR (elementary band representation) for the bands forming FS. Recently it was shown that all bands can be grouped into sets that correspond to a single EBR; topological semimetal behavior can be understood as a property of a partially occupied set of such bands. Also, the non-quantized contribution to AHE, as shown by Haldane et al\cite{haldane_berry_2004}, is expected to be a pure Fermi surface property. Combining these two ideas, a part of the FS that is comprised of multiple pockets created by the bands belonging to a single EBR, can be considered distinctly from \textit{another part} of the FS similarly corresponding to the \textit{bands from another EBR}. 

\indent{}In the case of a true semimetal, where there is a continuous gap disconnecting sets of bands contributing to the FS, $\mathbb{H}_F$ can be calculated separately using formula (8) for parts of the surfaces arising from distinct sets of bands (bands with differing EBRs), and subsequently summed together in order to characterize the entire FS. This is exactly the case for Co$_2$MnAl, CrPt$_3$ and KV$_3$Sb$_5$. For the case of KV$_3$Sb$_5$, it can be seen that the contributions of the distinct EBRs are not cooperative, resulting in a relatively low $\mathbb{H}_F$ of ~0.14. However, for Co$_2$MnAl and CrPt$_3$, both have cooperative contributions and correspondingly have H values larger than 1 as well as AHC values larger than 1570 $\frac{\hbar}{e}(\Omega cm)^{-1}$; but they still correlate extremely well with the overall trend in Figure 3c. Figure 4a showcases the detailed bandstructure for CrPt$_3$ with each distinct set of bands colored (blue and yellow) with the continuous gap shaded in gray. The insets clarify the almost-degeneracies near Gamma which are actually gapped. In the Berry curvature approach, the states from the different EBRs are mixed in the total calculation in the Kubo formula (4). Figure 4b shows the energy dependent AHC calculated from the Kubo formalism as well as the energy dependent AHC (using the AHC vs H correlation $\sigma = m\mathbb{H}_F + \sigma_0$ to convert $\mathbb{H}_F$ to a numerical AHC value). The results from the two methods are qualitatively similar, but the $\mathbb{H}_F$ result has a slightly better quantitative match to experiment. 

%about simplisity and cheapest of DFT for H 
\indent{}When compared with the current Berry curvature driven method for AHE/SHE prediction, the $\mathbb{H}_F$ index is, computationally, a much simpler metric as it requires just basic DFT calculation without Wannier projection, and thus can carried out at a significant reduction in time and cost. Importantly, this analysis method can easily be fully automated and implemented into material databases, and can enable artificial intelligence and machine learning based searches of large repositories of compounds for materials with desirable traits for technological applications. For now the $\mathbb{H}_F$ index is still a relatively rough estimation; it does not separate the effect of locally open FSPs vs irregular FSPs around hyperbolic points, shown in the Figure 1, and is limited to the cases of 2D AHE/SHEs. However, the numerical correlation of the AHE/SHE with $\mathbb{H}_F$ of $R^2$ $=$ 0.97 proves that the concept of using geometric classification of electronic structure manifolds is not just a "`blue-sky"' theoretical research effort; it has immediate applications to outstanding questions in condensed matter physics. The results may extend to the anomalous Nernst effect and the non-linear Hall effect as well, due to their similarity in origin to the AHE/SHE.  

\indent{}The concepts outlined here will alter the current paradigm of understanding the non-trivial transport regimes (like AHE/SHE) moving it to include geometrical properties of the band structure and FS, rather than just topological properties of the eigenstates of the Hamiltonian. This way semi-classical transport equations can be developed using the idea of a quasi-particle moving along a trajectory described by the geodesic equation of the curved energy-momentum manifold (a Riemannian manifold), rather then a free particle moving with group velocity $v_F$, \textit{in direct analogy} to the theory of general relativity for a traveler moving along the geodesics of a curved space-time manifold (also a Riemannian manifold). See the supplement for visual demonstration of this analogy (figure S5). Describing the smooth deformation of the Riemannian metric for the manifold (e.g. geodesic flow) is one route to connecting the Thurston geometries to quasi-particle dynamics and transport phenomena. An immediate direction of future work is to attempt re-conceptualize a particle's spin as a geometrical construction of symplectic form on the energy-momentum manifold, which is a natural property of any odd dimensional manifolds. This may lead to fundamental understanding of other exotic effects, but is beyond the scope of this work, which is introducing the use of modern geometrical methods to the electronic structure theory. Some of the important open questions stemming from these ideas are: can the other geometrical invariants aside from the Gaussian curvature be applied to identification of the transport properties in crystals? Can a full derivation of the FSGE on quasi-particles in reciprocal space be translated into real-space electron transport equations through use of Born reciprocity relations? Are there other obvious correlations between geometrical classes of FS regions and other non-fermi liquid transport phenomena (e.g. Euclidean and, say, electron correlation)? How do boundaries between FS regions interact and do they result into effectively turbulent quasi-particle dynamics on an energy-momentum manifold, and in what limits? What are the consequences of this in real-space?

\indent{}In summary, motivated by Perlmann, Hamilton, and Thurston's works, we have introduced the general concept of using modern \textit{geometrical classification} of multi-dimensional manifolds to characterize electronic structure manifolds and predict non-trivial transport phenomena. For now, we simplified the problem from 3D band structures to 2D Fermi surfaces and outlined the Fermi Surface Geometry Effect, through the use of tangent bundles and Gaussian curvature, that relates the hyperbolicity of the Fermi surface with anomalous and spin Hall effects. This concept has been applied to develop an index, $\mathbb{H}_F$, for describing the the "hyperbolicity' of the FS and showed a universal correlation (R$^2$ = 0.97) with experimentally measured intrinsic AHE values for 16 different compounds spanning a wide variety of crystal, chemical, and electronic structure families. An apparent maximum value, at $\mathbb{H}_F$ = 1, of ~1570 $\frac{\hbar}{e}(\Omega cm)^{-1}$ was determined for materials with an FS created by bands belonging to a single EBR; materials with multi-EBR FS's can, and do, break this limit as evidenced by CrPt$_3$ and Co$_2$MnAl. Use of the $\mathbb{H}_F$ index allows direct calculation of the AHE/SHE at a much lower computational cost than current methods by eliminating the need for Wannier projection and can be implemented with existing high throughput DFT methods and databases. This work highlights the importance of, and opportunities laying ahead for, developing a complete theory of \textit{geometrical understanding} of electronic structure manifolds beginning with Fermi surfaces. Also, these ideas can be extended to phononic and magnonic band structures as well. In analogy to the broad impact that topological understanding of these structures had, a theory of the Fermi Surface Geometry Effect and eventual extension to other dimensional manifolds, will lead to a deeper understanding of at least electron transport and have far reaching consequences in condensed matter physics.

\newpage
Methods:
\indent{}Our calculations have been performed by using the density-functional theory (DFT) with localized atomic orbital basis and the full potential as implemented in the code of full-potential local-orbital (FPLO)~\cite{Koepernik1999}. The exchange and correlation energy was considered in the generalized gradient approximation (GGA) level~\cite{perdew1996}. The electronic band structures were further confirmed by the calculations from $ab-initio$ code of \textsc{wien2k}~\cite{blaha2001}. In all the calculations we have used the experimentally measured lattice structures. For the calculation of FS we used 30x30x30 k-meshes, which was found to be a sufficiently dense k-mesh with reasonable computing time (by analysis of the $\mathbb{H}_F$ dependence of k-mesh density for Fe and Co$_2$FeSi; see Figure S2 for details).
\newpage

\begin{figure}[h]
	\includegraphics[width=1.0\textwidth]{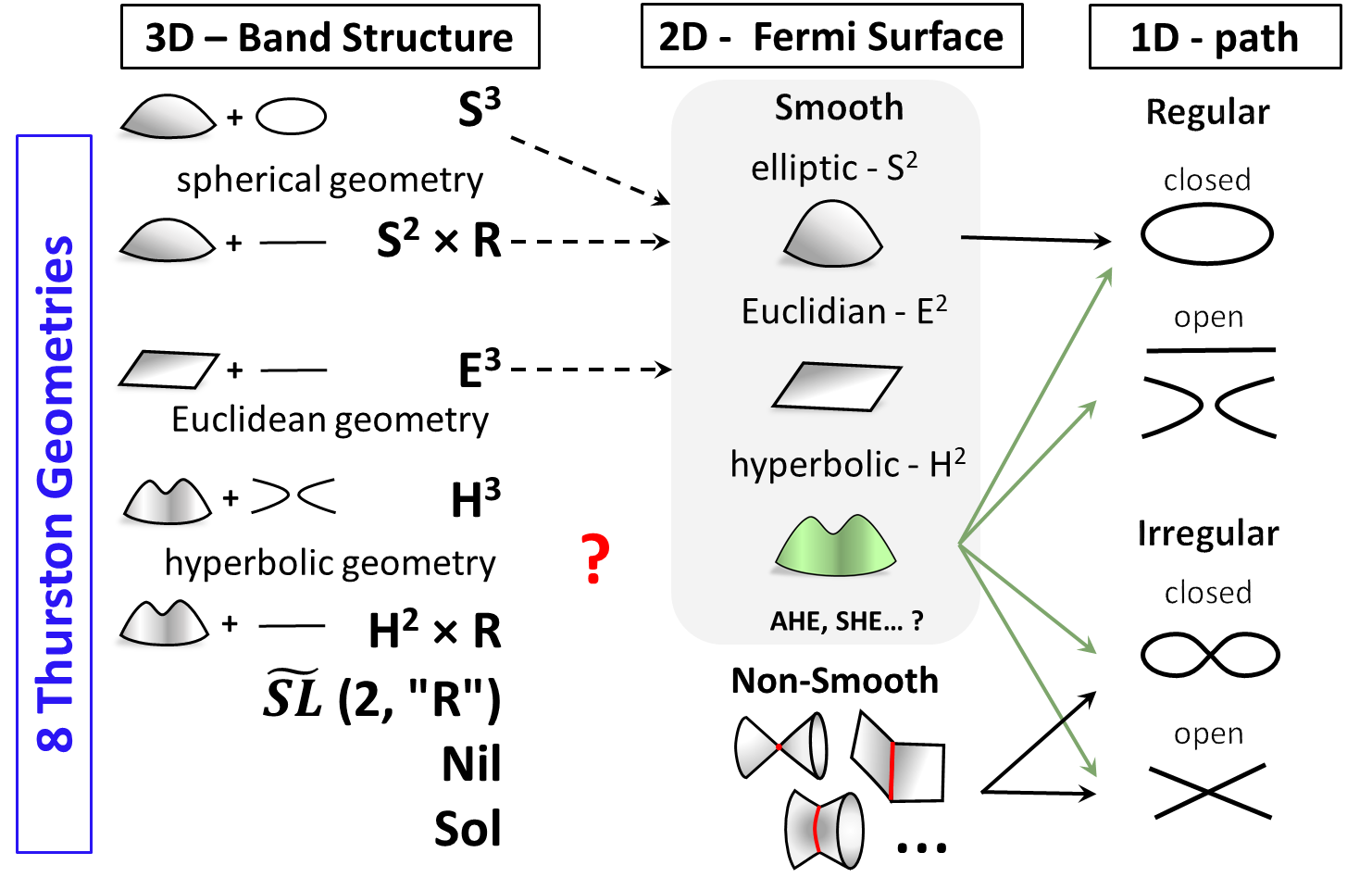}
	\caption{\scriptsize{\textbf{(color online):}Diagram of the classes of local geometries in different dimensions. The left column shows the 8 Thurston geometries for 3D manifolds, like band structures, presented schematically and with their symbols of geometry. The middle column shows the three smooth (and non-smooth) classes for 2D manifolds like Fermi surfaces. The right hand column shows the classifications of 1D paths, like Fermi surface paths, resulting from further reduction of dimension. Arrows illustrate the connection between certain higher dimension geometrical classes and their lower dimensional counterparts; green arrows showcase how hyperbolic 2D manifolds can result in all types of 1D paths.}}
	\label{Figure_1}
\end{figure}

\newpage

\begin{figure}[h]
	\includegraphics[width=1.0\textwidth]{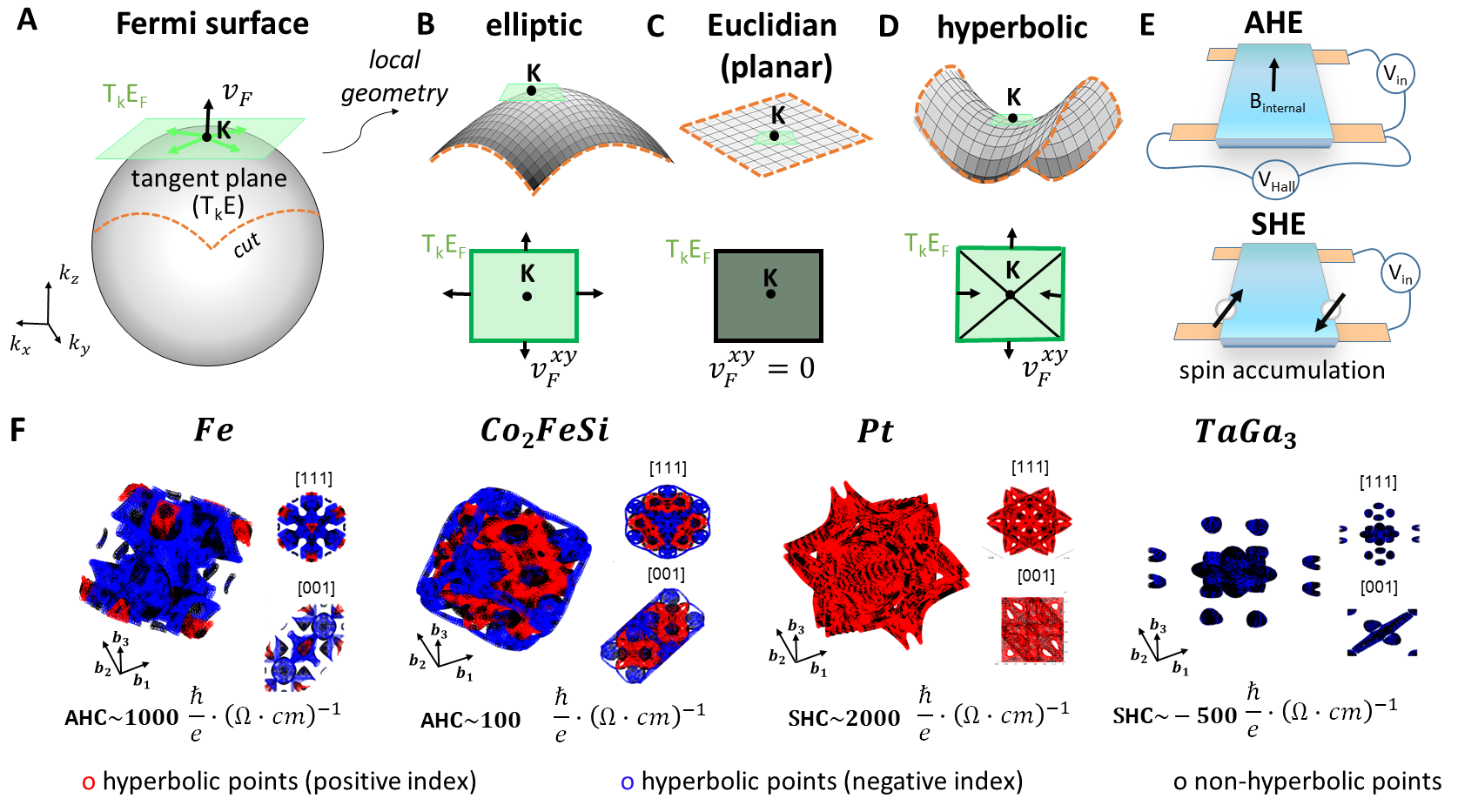}
	\caption{\scriptsize{\textbf{(color online):}A. Schematic representation a typical Fermi surface where the green plane (T$_K$E$_F$) represents the tangent plane at point K on the Fermi surface and v$_F$ is the group velocity. B: A locally elliptic cut of a Fermi surface with tangent plane and velocity vectors drawn as green arrows. C: A locally Euclidian cut and D: A locally hyperbolic cut. E. Illustration of  anomalous Hall (AHE)  and spin Hall (SHE) measurement geometries. F. Schematic distribution of the hyperbolic and non-hyperbolic regions of the Fermi surface in the reciprocal unit cell for several AHE and SHE compounds of varying magnitude.}}
	\label{Figure_2}
\end{figure}

\newpage

\begin{figure}[h]
	\includegraphics[width=1.0\textwidth]{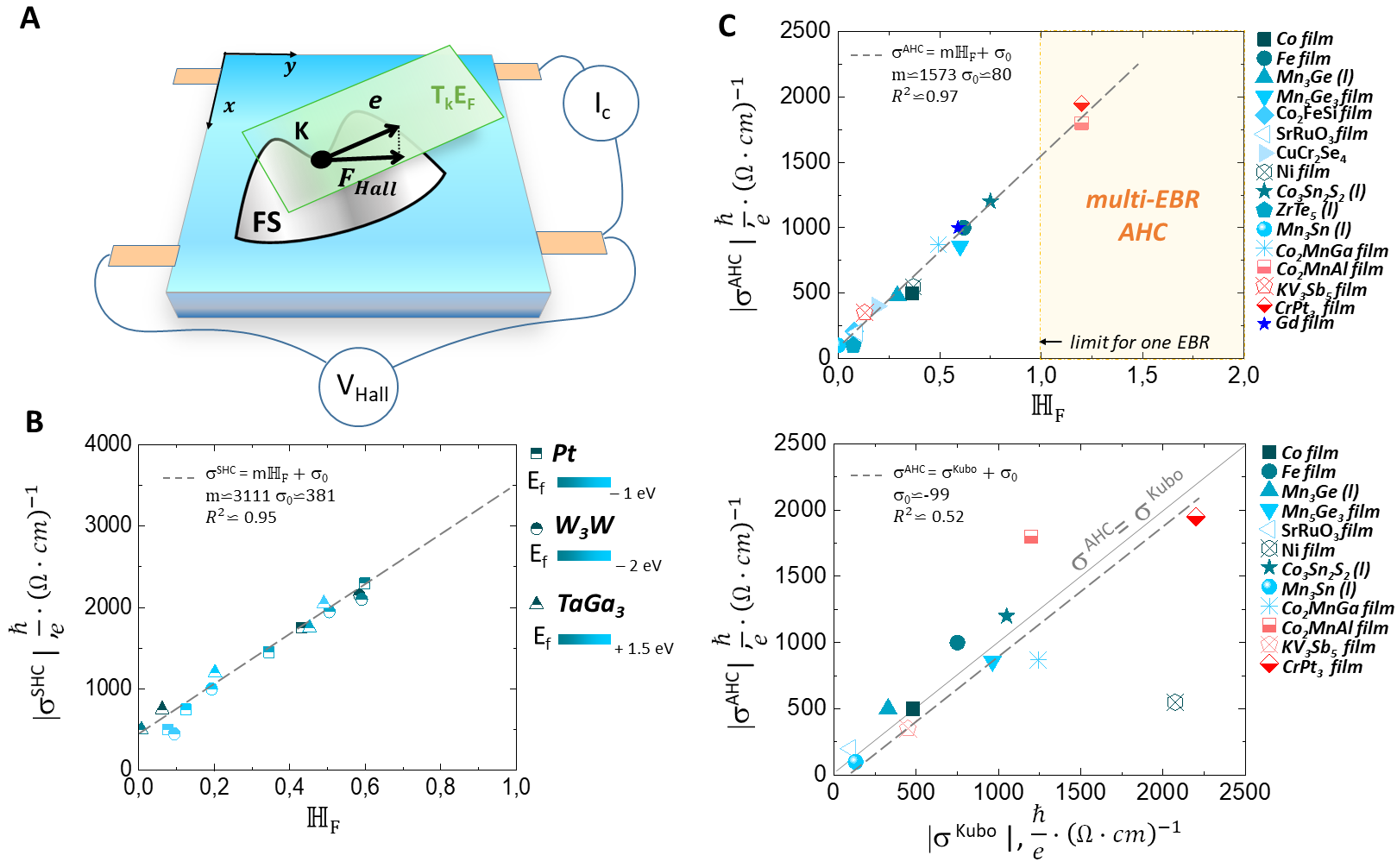}
	\caption{\scriptsize{\textbf{(color online):}A. Schematic image of the FSGE in the direction of Hall measurement. B. Correlation graph of the predicted SHC values via the Kubo formalism vs $\mathbb{H}_F$, as defined in the text. C. Correlation graph top: \textit{experimentally} determined intrinsic AHC vs $\mathbb{H}_F$ for various materials ((l) identifies layered structures). Bottom: \textit{experimentally} determined intrinsic AHC vs predicted values of AHC via the Kubo formalism.}}
	\label{Figure_3}
\end{figure}  

\newpage

\begin{figure}[h]
	\includegraphics[width=1.0\textwidth]{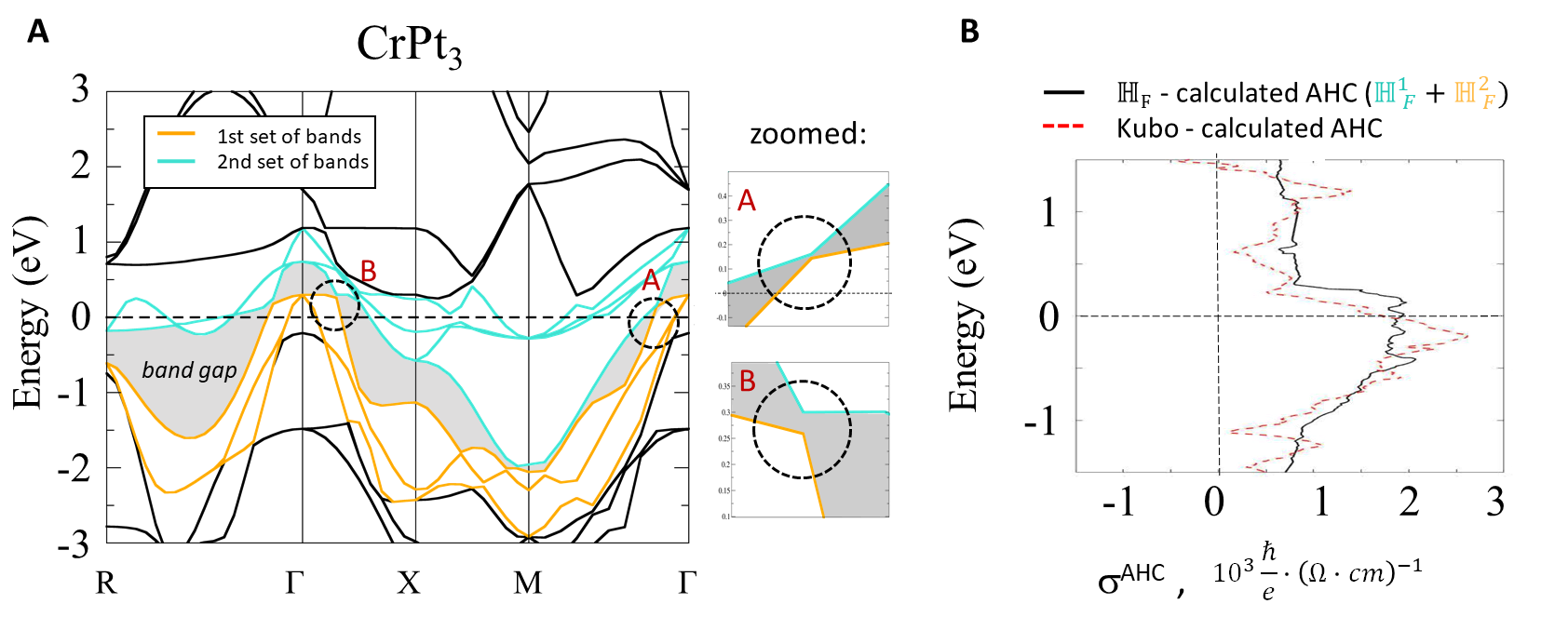}
	\caption{\scriptsize{\textbf{(color online):}A. Bandstrucure of $CrPt_3$. Blue and yellow colors represent two topologically disconnected (having different EBRs) sets of bands crossing the Fermi level. These sets are disconnected by the continuous gap present between them; i.e. true semimetallic behavior. B. Graph of energy resolved AHC predicted in two different ways: red dashed line is the Kubo based prediction, black dashed line stems from the linear correlation between $H_F$ and AHC calculated separately for FS contributions from each set of bands, then summed together for total $\mathbb{H}_F$.}}
	\label{Figure 4}
\end{figure}

\newpage
\bibliography{Lit}
\newpage
Acknowledgments: This research was supported by the Alexander von Humboldt Foundation Sofia Kovalevskaja Award and the BMBF MINERVA ARCHES Award. M.W.G. thanks the Leverhulme Trust for funding via the Leverhulme Research Centre for Functional Materials Design. All data needed to evaluate the conclusions in the paper are present in the paper and the Supplementary Materials.

Author Contributions: E.D. was the lead researcher on this project. She carried out the main theoretical derivations as well as the majority of the DFT calculations and analysis. M.G. and J.G. assisted with calculations and interpretation. M.N.A is the principal investigator.
 
Competing Interests The authors declare that they have no competing interests.

Correspondence Correspondence and requests for materials should be addressed to Mazhar N. Ali~(email: maz@berkeley.edu).
\end{document}